\documentclass{iaus}
\usepackage{graphicx}

\title[Emergent Exoplanet Flux] 
{Emergent Exoplanet Flux: \\ Review of the Spitzer Results}

\author{Drake Deming}   

\affiliation{Planetary Systems Laboratory \\ Code 693,
NASA's Goddard Space Flight Center \\
Greenbelt MD 20771 USA\\ email: {\tt ddeming@pop600.gsfc.nasa.gov}}

\pubyear{2008}
\volume{253}  
\pagerange{1--11}
\jname{Transiting Planets}
\editors{F.~Pont et al., eds.}
\begin{document}

\maketitle

\begin{abstract}
Observations using the {\it Spitzer Space Telescope} provided the
first detections of photons from extrasolar planets. {\it Spitzer}
observations are allowing us to infer the temperature structure,
composition, and dynamics of exoplanet atmospheres.  The {Spitzer}
studies extend from many hot Jupiters, to the hot Neptune orbiting
GJ\,436. Here I review the current status of {\it Spitzer} secondary
eclipse observations, and summarize the results from the viewpoint of
what is robust, what needs more work, and what the observations are
telling us about the physical nature of exoplanet atmospheres.
\end{abstract}

\firstsection 
\section{Introduction}

The powerful astrophysical leverage provided by transits enables us to
study extrasolar planets directly, i.e., by detection of their
emergent radiation.  The {\it Spitzer Space Telescope} has provided
the bulk of these detections.  The first {\it Spitzer} measurements of
exoplanet secondary eclipses were announced in 2005.  Two independent
groups (Charbonneau et al. 2005; Deming et al. 2005) measured eclipses
for two different planets, using two different {\it Spitzer}
instruments, and obtained very similar results (Figure~1).

\begin{figure}[ht]
\begin{center}
\includegraphics[width=2.8in]{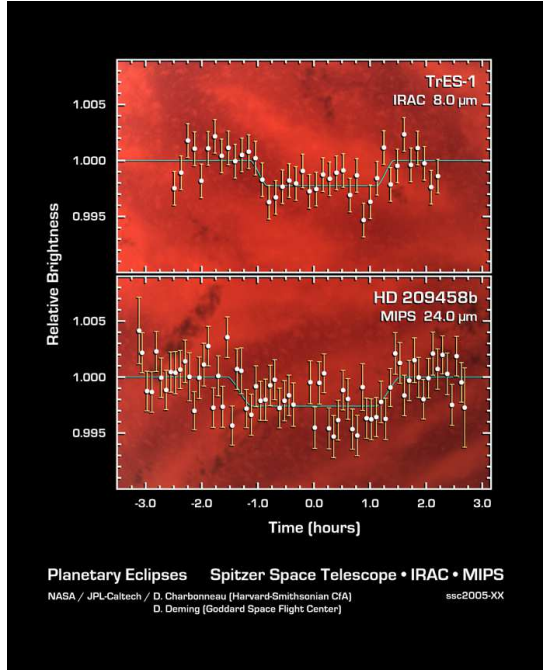}
 \caption{First detections of exoplanet thermal emission using the
 {\it Spitzer Space Telescope}. Plotted are the secondary eclipses of
 TrES-1 at 8\,$\mu$m (top, Charbonneau et al. 2005), and HD\,209458b
 at 24\,$\mu$m (bottom, Deming et al. 2005)}.
   \label{fig1}
\end{center}
\end{figure}

Since each eclipse was independently measured to $\sim 6\sigma$
significance, exoplanet thermal emission was securely detected.  The
discovery of transits in HD\,189733b (Bouchy et al. 2005) provided an
opportunity to measure exoplanet thermal emission at higher 
signal-to-noise ratio.  Initial observations of HD\,189733b at
16\,$\mu$m (Deming et al. 2006) showed an eclipse of the planet at
32$\sigma$ significance (Figure~2), and subsequent work using the IRAC
instrument has detetcted the planet's flux to 60$\sigma$ precision at
8\,$\mu$m (Knutson et al. 2007).  This extraordinary level of
precision in measuring exoplanet thermal emission allows many
intersting studies that could hardly have been imagined when the first
extrasolar planets were detected by radial velocity studies.

\begin{figure}[ht]
\begin{center}
\includegraphics[width=2.2in]{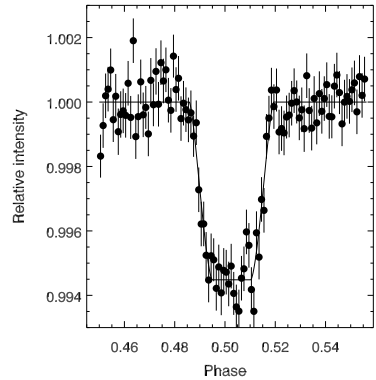}
 \caption{Eclipse of HD\,189733b at 16\,$\mu$m (Deming et al. 2006)}.
   \label{fig2}
\end{center}
\end{figure}

In this review I summarize highlights from {\it Spitzer} secondary
eclipse measurements, with some discussion of transmission
spectroscopy during transit. The quality of work in this field has
been uniformly high, but I will summarize the results from the
viewpoint of what is robust, and what I believe needs more work and
clarification.

\section{Spectral Energy Distributions from Photometry}

The depth of a planetary eclipse measures the brightness of the planet
at that wavelength, in units of the stellar brightness.  Combining
results from multiple eclipses at different wavelengths allows us to
reconstruct the spectral energy distribution of the planet at
photometric spectral resolution (typically, $\lambda/\delta \lambda
\sim 3$).  The results are usually shown in `contrast' units, i.e.,
planet divided by star, since that is what we actually measure. The
contrast amplitude of exoplanet eclipses is greatest at the longest
wavelengths.  In the Rayleigh-Jeans limit, the eclipse amplitude
($A_{\lambda}$) is (Charbonneau 2003):
\begin{equation}
A_{\lambda} = (R_p^2/R_s^2)(T_p/T_s),
\end{equation}

\noindent where $(R_p^2/R_s^2)$ is the ratio of planet-to-star area, and
$(T_p/T_s)$ is the ratio of planet-to-star brightness temperature.
Assuming that the planet and star resemble blackbodies (a reasonable
approximation at the longest wavelengths), then:
\begin{equation}
T_p =\alpha T_s \theta^{1/2},
\end{equation}

\noindent where $\theta$ is the angular diameter of the star as seen from the
planet, and $\alpha$ is a constant that contains the planet's albedo,
circulation properties, etc.  An important corollary of Eqs. 2.1 and
2.2 is the effect of smaller parent stars (e.g., M-dwarfs).  As we
proceed down the main sequence, $R_s$ and $\theta$ decrease, but the
exponents imply that the $(R_p^2/R_s^2)$ term dominates over
$\theta^{1/2}$.  Hence planets orbiting small stars will generally
exhibit deeper eclipses, and their emergent flux will be more
detectable for that reason. Moreover, the habitable zone moves closer
to lower main sequence stars, and the transit probability increases
inversely as the planet's orbit radius. This circumstance is a major impetus for
finding terrestrial planets in the habitable zones of M-dwarfs
(Charbonneau and Deming, 2007).

The time of the secondary eclipse is very sensitive to the
eccentricity of the orbit, specifically to $e\,{\cos}\,\omega$
(Charbonneau 2003).  For example, the eclipse of GJ\,436b occurs at
phase $0.585\pm0.005$ (Deming et al. 2007), more than five hours after
the mid-point between transits.  Since the eclipse duration is
$\sim$1-hour, the sensitivity of the eclipse time to moderately small
eccentricity ($e=0.15$ for GJ\,436b) is obvious.  


\subsection{Molecular Absorption}

The actual flux from close-in planets will peak near 2 to 5\,$\mu$m,
not at the wavelengths of greatest contrast.  Moreover, the shorter
infrared (IR) wavelengths are key to inferring the composition and
temperature structure of the planet's atmosphere.  Figure~3 shows the
spectrum of HD\,189733b, in flux (not contrast) units, from Barman
(2008). The IR spectra of hot Jupiters are believed to be shaped
predominantly by water absorption (Burrows et al. 2005, Seager et
al. 2005), but other molecules such as methane also play a role (e.g.,
Swain et al. 2008a), and methane in particular could become more
important for cooler planets like GJ\,436b. For close-in planets
orbiting luminous stars, strong irradiation could flatten the
temperature gradient and weaken absorption features in the spectrum at
the time of eclipse (Fortney et al. 2006). {\it Spitzer} results from
spectroscopy initially suggested that water absorption might not be a
prominent feature in eclipse spectra (Grillmair et al. 2007,
Richardson et al 2007, see discussion below). However, the HD\,189733b
results from IRAC (Charbonneau et al. 2008) are in good accord with
Barman's standard model, and provide convincing evidence that water
absorption shapes the 2- to 5\,$\mu$m spectra of at least one hot
Jupiter.

Important observations of {\it transits} have also been made using
{\it Spitzer} (Richardson et al. 2006, Gillon et al. 2007, Nutzman et
al. 2008, Agol et al. 2008), including evidence for water absorption
during transit (Tinetti et al. 2007, Beaulieu et al. 2008).
Ehrenreich et al. (2007) have suggested that better correction for
instrument systematics is needed before we can conclude that water
absorption is detected via IRAC photometry during transit.  While I
believe there is good evidence for water absorption in transit,
additional work to better understand {\it Spitzer's} instrumental
systematics is certainly warranted.

\begin{figure}[ht]
\begin{center}
\includegraphics[width=2.5in]{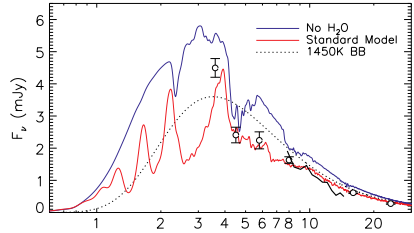}
 \caption{Modeled spectrum of HD\,189733b, giving flux vs. wavelength
 in microns, from Barman (2008), with measurements from Charbonneau et
 al. (2008).  The standard model including water absorption provides
 the best fit to the measurements.}
   \label{fig3}
\end{center}
\end{figure}

\subsection{Circulation and Dynamics}

Hot Jupiters are believed to rotate synchronously with their orbital
period, keeping one side perpetually pointed toward the star.  An
important question is the degree to which they transport heat to the
anti-stellar hemisphere via strong atmospheric circulation.  Models of
this circulation (e.g., Showman \& Guillot 2002, Cho et al. 2003, 2008, 
Cooper and Showman 2006, Langton and Laughlin 2008) can be checked
using {\it Spitzer} around-the-orbit observations, either made
continuously (Knutson et al. 2007), or via periodic sampling
(Harrington et al. 2006, Cowan et al. 2007). Observations of two
non-transiting hot Jupiters, Ups And\,b at 24\,$\mu$m (Harrington et
al. 2006) and HD\,179949b at 8\,$\mu$m (Cowan et al. 2007) suggested
large day-to-night temperature contrasts.  Other planets give much
lower day-night contrast (Knutson et al. 2007, Cowan et al. 2007). One
possibility is that the difference for Ups And\,b is related to the
greater formation height of 24\,$\mu$m radiation. But note that
Knutson et al. (2008a) have observed HD\,189733b at 24\,$\mu$m, and
find a result commensurate with their 8\,$\mu$m results.  Another
possibility is that these type of phased observations could be
affected by a temporary `hot spot', and would not typically show such
a large day-night difference. However, in that case we would also
expect greater-than-predicted variability at secondary eclipse for
some transiting planets, and even low-level variability has not yet
been observed. {\it Spitzer} will re-observe Ups And\,b at 24\,$\mu$m
(B. Hansen, private communication) if the cryogen lasts long enough.

In addition to close-in planets on circular orbits, {\it Spitzer} has
great potential to observe the time-dependent heating (Iro \& Deming
2008) for planets on very elliptic orbits.  Recently, Laughlin et
al. (2008) observed the flash-heating of HD\,80606b at periastron,
and {\it Spitzer} may be able to make more observations of this type
during the warm mission.

\subsection{Inverted Temperature Gradients}

Although HD\,189733b shows water absorption, and agrees well with
standard models, there is evidence that other exoplanets show
different atmospheric structures.  HD\,209458b exhibits an atmospheric
temperature inversion (i.e., temperature rises with increasing
height). The first hint of this inversion was found by Richardson et
al. (2007), who derived emission features in their eclipse spectrum at
relatively `high' spectral resolution $\lambda/\delta\lambda \sim
100$, albeit at low signal-to-noise.  Definitive evidence of the
inversion comes from the IRAC eclipse measurements by Knutson et
al. (2008b).  Figure~4 shows the IRAC measurements for both
HD\,189733b (Charbonneau et al. 2008, Barman 2008) and HD\,209458b
(Knutson et al. 2008b, Burrows et al. 2007), compared with models from
Burrows et al. (2008).  In comparing the observations and models on
Figure~4, I've scaled the models by an arbitrary contrast factor to
produce the best fit by eye.  This serves to illustrate the nature of
these two different exoplanet spectra, but I caution that for the
quantitative fits, readers should consult the original papers.

\begin{figure}[ht]
\begin{center}
\includegraphics[width=3.0in]{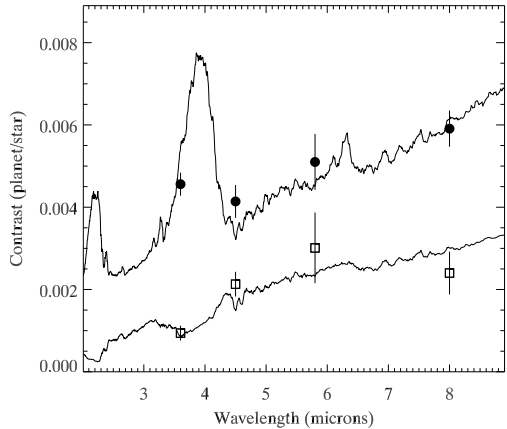}
 \caption{Measurements of HD\,189733b from Charbonneau et al. (2008),
 and HD\,209458b from Knutson et al. (2008b), compared to a standard
 model (HD\,189733b) and a temperature-inverted model (HD\,209458b)
 from Burrows et al. (2008). The model and observations for
 HD\,189733b have been offset upward by 0.002 for clarity. Error bars are $\pm 2\sigma$.}
   \label{fig4}
\end{center}
\end{figure}

In addition to IRAC, {\it Spitzer} eclipse observations of both
planets have been obtained at 16\,$\mu$m and 24\,$\mu$m (not plotted).
These longer wavelengths are also sensitive to water absorption and
atmospheric temperature gradients, but less so than at IRAC
wavelengths. When comparing spectral energy distributions for
different planets, remember that all four IRAC wavelengths were observed
simultaneously only for HD\,189733b and HD\,209458b.  More commonly,
due to the nature of the instrument (Fazio et al. 2005), eclipses are
measured at 3.6 and 5.8\,$\mu$m simultaneously, and then in another
eclipse at 4.5 and 8\,$\mu$m simultaneously. In principle, variability
of the planet's thermal emission (Rauscher et al. 2007) could
contaminate the measured spectrum.  To date, there is no evidence for
variability large enough to produce significant spectrum errors (see
Agol et al. 2008).

The hallmark of an inverted spectrum can be seen by comparing 5.8 and
8.0\,$\mu$m measurements, as well as comparing 3.6 and
4.5\,$\mu$m. The inverted atmosphere has a higher 5.8\,$\mu$m flux
than 8.0\,$\mu$m, because 5.8\,$\mu$m has high opacity due to water
vapor, and the water bands are present in emission. In the
non-inverted atmosphere the 3.6\,$\mu$m flux is elevated because
the lesser opacity at this wavelength allows planet flux to well up
from the deeper atmosphere, where temperatures are higher.  The
non-inverted 4.5\,$\mu$m band shows lower flux because water and CO
opacity cause the radiation to be emitted from higher layers of the
atmosphere, where the temperature is lower.  The inverted atmosphere
may have a high altitude absorbing layer (Burrows et al. 2008), and
this raises the temperature at the 4.5\,$\mu$m height, and lowers it
at the 3.6\,$\mu$m height, reversing the relative magnitudes of the
emergent fluxes.

An open question is the physical cause of the inversion, and whether
this is a common phenomenon in hot Jupiter atmospheres.  Burrows et
al. (2008) attribute the inversion to the presence of a high altitude
optical absorbing layer, but the composition and origin of this layer
are unknown.  Doubtless the high level of stellar irradiation that hot
Jupiters experience plays a major role in inverting the temperature
gradient.  Fortney et al. (2008) define two classes of hot Jupiters
depending on whether the stellar irradiation drives the formation of a
hot stratosphere (i.e., region of higher temperature) via TiO/VO
absorption. TiO and VO have bands in the optical where stellar fluxes
are high, and have been implicated in perturbations to exoplanet
atmospheric temperature structure (Hubeny et al. 2003).  Figure~5
shows many of the known exoplanets in mass vs. irradiance
space, with the predicted boundary between the pM class
(stratospheres) and pL class (no stratospheres) indicated. I have
circled the three planets that are known or strongly suspected to have
inverted temperature gradients.  Besides HD\,209458b, HD\,149026b is
believed to have a hot stratosphere on the basis of the very high
brightness temperature at 8\,$\mu$m (Harrington et al. 2007).
Recently, XO-1b was found to have an inverted gradient (Machalek et
al. 2008).  Although XO-1b is predicted to be pL class, it is at the
upper range of that class, so there is no firm counterexample to the
Fortney et al. (2008) theory at this point.  During this conference,
new information is becoming available on TrES-2 (O'Donovan et
al. 2008) and TrES-4 (Knutson et al. 2008c), and {\it Spitzer} data
for several other planets are under analysis.  Hence we should soon
learn how well the curent pM/pL classification theory corresponds to reality.

\begin{figure}[ht]
\begin{center}
\includegraphics[width=2.5in]{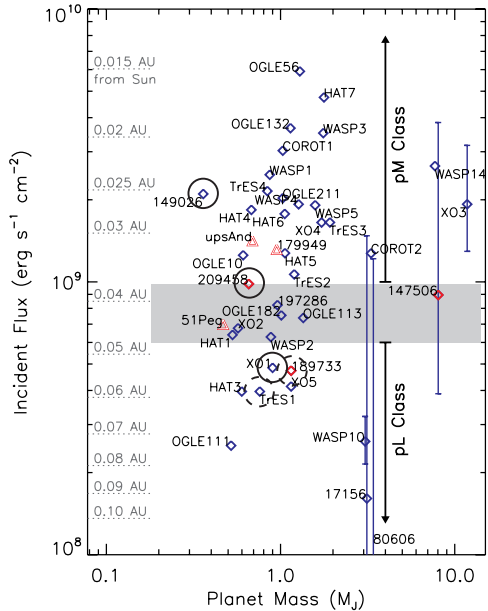}
 \caption{Atmospheric structure classification for hot Jupiters, from
 Fortney et al. (2008).  The pM class planets are predicted to have
 hot stratospheres (i.e., temperature inversions), whereas the pL
 class planets should not. The planets that are currently known or
 strongly suspected to have inversions are circled with solid lines,
 and the non-inverted planets (TrES-1, and HD\,189733b) are circled with dashed
 lines.}
   \label{fig5}
\end{center}
\end{figure}

\subsection{A Hot Neptune}

{\it Spitzer} secondary eclipse observations extend down to the hot
Neptune orbiting GJ\,436 (Deming et al. 2007a, Demory et al. 2007).
Figure~6 shows this eclipse at 8\,$\mu$m.  The inferred brightness
temperature ($712\pm36$K) is modestly above the predicted brightness
temperature for thermal equilibrium with the star (Deming et
al. 2007a), but the uncertainties are relatively large.  Additional
{\it Spitzer} eclipse observations were recently made at at 8- and
24\,$\mu$m (J. Harrington, private communication) and
`around-the-orbit' observations (by Knutson et al.) are pending.  The
totality of {\it Spitzer} observations may be sufficient to define the
total luminosity of the planet, and thereby determine whether it emits
significant energy due to tidal dissipation in its moderately
eccentric orbit.

\begin{figure}[ht]
\begin{center}
\includegraphics[width=3.0in]{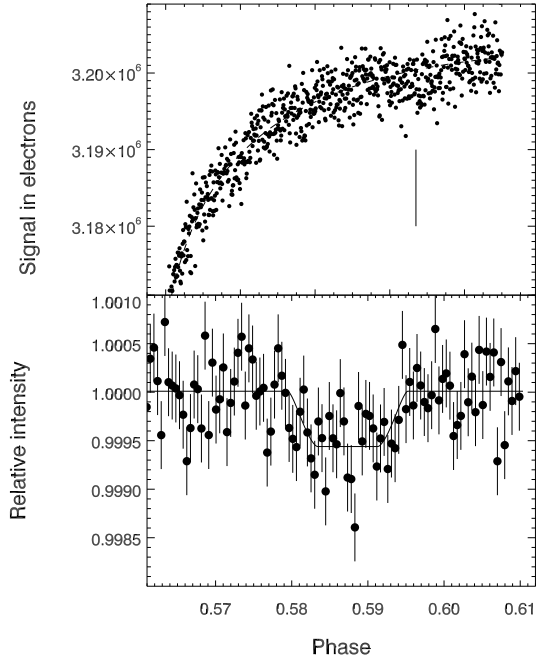}
 \caption{Secondary eclipse of the hot Neptune planet GJ\,436b
 observed by {\it Spitzer} at 8\,$\mu$m (Deming et al. 2007a). The top
 panel shows the unbinned data prior to correction of the detector
 ramp, and the lower panel shows binned data and a fit of an eclipse
 curve centered at phase 0.587, indicative of an eccentric orbit.}
   \label{fig6}
\end{center}
\end{figure}

\section{Spectroscopy}

{\it Spitzer} secondary eclipse observations have been extended to
spectroscopy as well as photometry (Grillmair et al. 2007, Richardson
et al. 2007, Swain et al. 2008b).  The principle of these measurements
is simple.  Suppose that an absorption (emission) feature occurs in
the planets atmosphere at a given wavelength.  Then the depth of the
eclipse at that wavelength will be smaller (larger) than at other
wavelengths.  Hence the emergent spectrum of the planet can be
constructed from the wavelength dependence of the eclipse depth.  In
practice, this is a much more difficult observation than {\it Spitzer}
photometry, for two reasons.  First, there are many fewer photons per
wavelength channel because the light is dispersed, so the
signal-to-noise ratio is lower than for photometry.  Second,
spectroscopy is more affected by instrument systematic effects, as
discussed below.  In spite of these difficulties, the results are of
great interest.  Two exoplanets are sufficiently bright to make {\it
Spitzer} spectroscopy practical: HD\,189733b (Grillmair et al. 2007),
and HD\,209458b (Richardson et al. 2007, Swain et al. 2008b).  The
initial results indicated that these spectra were remarkably flat from
$\sim 7$ to $\sim 13\,\mu$m, not showing absorption due to water vapor
that was expected shortward of $\sim 8\,\mu$m.  However, additional
spectroscopy of HD\,189733b (Grillmair, private communication) agrees
very well with {\it Spitzer} photometry and with model atmosphere
predictions.  HD\,209468b having an inverted temperature gradient is
consistent with the flatness observed in the spectrum by Richardson et
al. (2007).

\begin{figure}[ht]
\begin{center}
\includegraphics[width=3.0in]{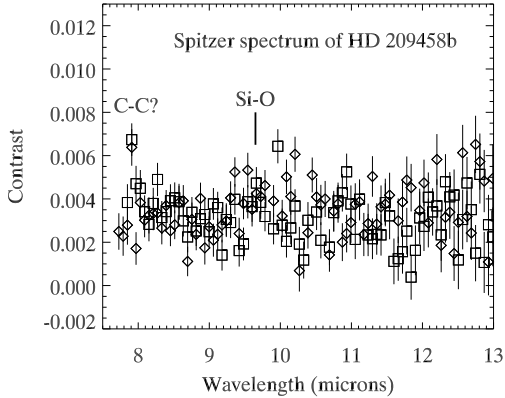}
 \caption{Spectrum of HD\,209458b, derived using an analysis very
 similar to Richardson et al. (2007), based on two eclipses (different
 symbols).  The two emission features evident in the spectrum are a
 broad bump near 9.8\,$\mu$m that Richardson et al. attribute to
 silicate clouds, and a sharp emission feature at 7.78\,$\mu$m
 possibly due to a C-C stretching resonance.}
   \label{fig7}
\end{center}
\end{figure}

Grillmair et al. (2007) did not find evidence for discrete spectral
absorption/emission features in the spectrum of HD\,189733b, but
Richardson et al. (2007) concluded that the spectrum of HD\,209458b
contained two discrete features, both present in emission.  Near
9.8\,$\mu$m they found evidence for a relatively broad emission bump
that they attributed to the Si-O stretching resonance, and at
7.78\,$\mu$m they find a sharp emission feature that may be due to a
C-C stretching resonance.  To illustrate these results, I've
re-analyzed the Richardson et al. data, using a similar method, and
this result is shown in Figure~7.  These data have also been analyzed
by Swain et al. (2008b), using an entirely different method.  Swain et
al. find emission near 7.8\,$\mu$m, but not at 9.8\,$\mu$m.  This
difference in results seems consistent with the nature of the
systematic effects in the IRS instrument: broad features (9.8\,$\mu$m)
are more sensitive to the way the instrument systematics are treated
in the analysis, whereas sharp features (7.8\,$\mu$m) are insensitive
to the analysis method.  It is important to optimize the spectroscopic
eclipse technique, so that we can use it with JWST to measure the
spectra of potentially habitable planets transiting M-dwarfs
(Charbonneau and Deming, 2007).

\section{{\it Spitzer} Instrument Systematics}

{\it Spitzer} observations have proven to be a remarkably stable and
sensitive way to measure exoplanet thermal emission. {\it Spitzer}
exoplanet aperture photometry (for $\lambda \leq 8\,\mu$m) achieves
noise levels closely approaching the photon noise limit, and the
errors average down as the inverse square root of exposure time.  For
the longer {\it Spitzer} wavelengths, where the zodiacal thermal
background is significant, the most precise photometry often requires
PSF-weighted optimal photometry, depending on the brightness of the
star. Unlike ground-based photometry, it is generally {\it not} necessary to
use `comparison stars' with {\it Spitzer}.  In fact, it can even be
detrimental to rely on comparison stars, because {\it Spitzer} does
have instrument systematic effects that could vary with position on
the detector. There are several effects that are currently recognized
and accounted for in {\it Spitzer} analyses, and most of them are now
described in the {\it Spitzer} instrument documentation.  The ones
most relevant to exoplanet eclipses are:

{\it The Ramp}.  Photometry at 8- and 16\,$\mu$m exhibits a gradually
increasing intensity, equivalent to an increasing gain in the
instrument response. This apparent gain increase is flux-dependent:
bright sources reach maxmium intensity more rapidly than faint
sources. This so-called `ramp' (Deming et al. 2006) is obvious in the
top panel of Figure~6.  Knutson et al. (2007) hypothesize that it is
due to charge-trapping, which is (so far) the most promising
hypothesis. Note that the ramp is {\it not} simply due to build-up of a
latent image, since none is present when the telescope is nodded
(Deming et al. 2006). In the charge-trapping hypothesis, the first
electrons generated by photons are captured by ionized impurites in
the detector material, and do not contribute to the signal on the
observed time scale.  As the detector is exposed to additional
radiation, the charge traps saturate, and the signal readout reaches
an asymptotic level.  This explanation is broadly consistent with the
known characteristics of the ramp, with some exceptions.  Observations
at 5.8\,$\mu$m can exhibit a `negative' ramp, i.e. a decreasing
intensity with time (Machalek et al. 2008), and no ramp is seen in
24\,$\mu$m photometry (Knutson and Charbonneau, private
communication). The existence of decreasing, as well as increasing,
ramps suggests a complex phenomenon that may depend on the gate and
bias voltages applied to the detector.  The lack of a ramp at
24\,$\mu$m may be due to the fact that the relatively large zodiacal
background at this wavelength keeps the ramp perpetually at its
maximum value, but the zodiacal background is also strong for
16\,$\mu$m photometry, which does exhibit a prominent ramp (Deming et
al. 2006). Spectroscopy using IRS also exhibits a ramp, at least when
obtaining spectroscopy at 7-14\,$\mu$m (Richardson et al. 2007).

The ramp is a relatively benign effect for eclipse photometry of
bright and high-contrast systems. The time scale for the ramp to reach
its maximum value is significantly longer than the duration of an
eclipse, so it's essentially a baseline effect that is included
when fitting to the eclipse depth.  But the ramp is more problematic
for fainter and low-contrast systems because small uncertainties in
the ramp curvature become significant relative to the eclipse
depth. The ramp is also problematic for `around-the-orbit'
observations where the planet signal will vary on a longer time
scale. One promising approach for this type of observation is to
`pre-flash' the detector by exposing it to a bright source immediately
prior to the exoplanet observations.  This saturates the ramp before
the exoplanet is observed; preliminary examination of observations
using a pre-flash (H. Knutson, private communication) suggest that the
technique is largely successful.

{\it Pixel Phase}. The pixels in the IRAC detectors are more
responsive when stellar images are centered on the pixels than when
they lie near the edges, and this is called the pixel phase effect.
One exception to this is very bright stars that are near
saturation. Detector non-linearity can produce a lower signal when
very bright stars are centered on a given pixel, but this circumstance
is normally avoided by using shorter integration times.  The pixel
phase effect is ubiquitous in the 3.6 and 4.5\,$\mu$m channels of
IRAC, and may be present to a much lesser degree at 5.8- and
8\,$\mu$m. Because there is pointing jitter in the telescope ($\sim$
tens of milli-arcsec), the pixel phase effect leads to a variable
intensity when performing aperture photometry. This is corrected in
eclipse data by decorrelating the intensity versus distance from pixel
center (e.g., Charbonneau et al. 2005, 2008).  Pixel phase is also
normally decorrelated from photometry performed for other {\it
Spitzer} research (Morales-Calderon et al. 2006), not just exoplanet
photometry.
 
{\it Spectroscopic Slit Losses}.  The IRS spectrometer (Houck et
al. 2005) has light losses at its entrance slit, like most
astronomical slit spectrometers.  Diffraction causes stellar images at
the slit to increase in size proportional to wavelength, so the slit
losses increase with wavelength also. Because there is telescope
pointing jitter on a time scale of $\sim1$\,hour, and possibly on
longer times scales, the intensity in a stellar spectrum will vary
slightly with both wavelength and time.  Even a slight variation in
the measured stellar spectrum distorts the spectra that are inferred
for exoplanets using the eclipse technique. There is no robust and
independent method to ascertain the exact telescope pointing for a
given spectrum.  Hence exoplanet observers correct for this distortion
in different ways (Richardson et al. 2007, Swain et
al. 2008b). Because the effect varies slowly with wavelength, it
primarily affects broad features in exoplanet spectra, not sharp
features like the 7.8\,$\mu$m emission inferred in HD\,209458b
(Richardson et al. 2007, and Figure 7). Hence the appearance of broad
features (like the 9.8\,$\mu$m peak on Figure 7), and indeed their
reality in exoplanet spectra, can depend on the method used to analyze
the data. Sharp features should be robust against instrument
systematics, but I note that the 7.8\,$\mu$m feature has relatively
low signal-to-noise.

\section{Warm Spitzer}

The depletion of cryogen (in $\sim$ April 2009) will terminate the
cold portion of the {\it Spitzer} mission.  However, the 3.6 and
4.5\,$\mu$m channels of IRAC will still operate at full sensitivity,
because the observatory will remain radiatively cooled at $\sim
35$K. The requirement to operate the `warm' mission at low cost
dictates that operations must be simple.  This favors relatively large
programs, and exoplanet transit and eclipse science is poised to take
full advantage of {\it Warm Spitzer}.  A `not-yet-obsolete' discussion
of some possible exoplanet applications is given in Deming et al. (2007b).

\section{Tabular Summary of the Spitzer Results}

The legacy of exoplanet science from cryogenic {\it Spitzer} has
already revolutionized exoplanet science, but has not yet reached full
fruition.  The results are of uniformly high quality, but the
observations are not easy.  The table in Figure~8 summarizes my
personal opinion as to which of the {\it Spitzer} results are already
robust, and which aspects need clarification by additional work.

\medskip
{\bf Acknowledgements.}  Adam Burrows, Travis Barman, Dave
Charbonneau, Jonathan Fortney, Heather Knutson, and Francis O'Donovan
all made valuable comments on a draft of this review.  I thank Travis
and Jonathan for providing Figures 3 and 5.

\begin{figure}[ht]
\begin{center}
\includegraphics[width=3.0in]{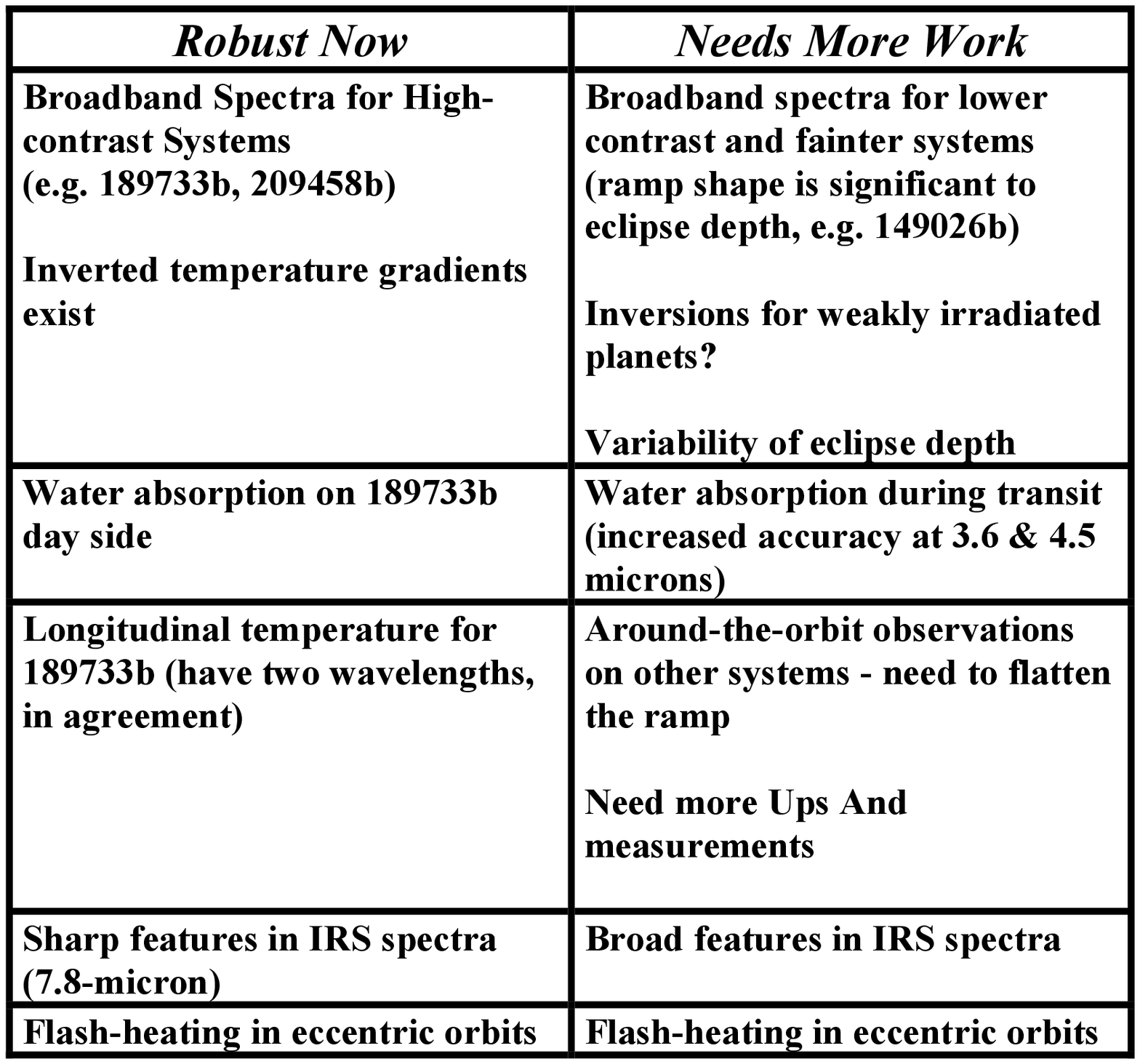}
 \caption{Tabular summary of the Spitzer results, categorized by
  results that are already robust, and those where more work is
  needed. Observations of eccentric planets fall into both categories,
  because the current work should be extended to more planets.}
   \label{fig8}
\end{center}
\end{figure}

\end{document}